\documentclass[pdftex,singlecolumn,epjc3]{svjour3} 
\RequirePackage{graphicx}
\RequirePackage{mathptmx}      
\RequirePackage{flushend}
\RequirePackage[numbers,sort&compress]{natbib}
\RequirePackage[colorlinks,citecolor=blue,urlcolor=blue,linkcolor=blue]{hyperref}
\usepackage{amsmath}
\usepackage{graphicx}
\usepackage{geometry}
\usepackage{caption}
\usepackage{subcaption}
\usepackage{color}
\begin{document}
\baselineskip=18 pt
\begin{center}
{\large{\bf Effects of Lorentz Symmetry Breaking Environment on Generalized Relativistic Quantum Oscillator Field }}
\end{center}

\vspace{0.5cm}

\begin{center}
{\bf Faizuddin Ahmed}\footnote{\bf faizuddinahmed15@gmail.com ; faizuddin@ustm.ac.in}\\
{\bf Department of Physics, University of Science \& Technology Meghalaya, Ri-Bhoi, Meghalaya-793101, India }
\end{center}

\vspace{0.5cm}

\begin{abstract}:
In this paper, we study the generalized Klein-Gordon oscillator under the effects of the violation of Lorentz Symmetry defined by a tensor field $(K_F)_{\mu\nu\alpha\beta}$ out of the Standard Model Extension (SME). We consider a possible scenario of Lorentz-Violating with a Cornell-type potential form electric field and a linear magnetic field that contributes a harmonic-type central potential effects in the quantum motions of oscillator fields. The bound-states solutions of the wave equation using the parametric Nikiforov-Uvarov method by considering a Coulomb- and Cornell-type potential form functions are obtained. We see that the eigenvalue solutions gets modified by the Lorentz symmetry breaking effects in comparison to the Landau levels (without Lorentz-Violation effects in flat space). 
\end{abstract}

\vspace{0.5cm}


\vspace{0.5cm}

{\bf PACS Number(s):} 11.30.Cp, 03.65.Pm, 03.65. Fd, 03.65.Ge,

\section{Introduction}

The Standard Model extension (SME) in high energy physics is an effective field theory that incorporates known physics and Lorentz-violating terms \cite{DC4,DC3,DC,DC2,DC0}. The violations enter through Lorentz non invariant operators in the Lagrangian, parameterized by coefficient tensors with Lorentz indices. If Lorentz symmetry is broken spontaneously, these coefficients are the vacuum expectation values of tensor operators, selecting out preferred directions in space-time. This SME is an extension of the usual Standard Model of particle physics including gravity along with all possible operators for Lorentz violation. The gauge sector of the standard model extension (SME) composed of a CPT-odd part \cite{b1} and a CPT-even part \cite{b1,b3,b4,b2,b6,b7,RC2}. Phenomenological consequences of Lorentz-violating and CPT-odd non minimal coupling between fermions and the gauge field in electromagnetism were first studied in Ref. \cite{b1} where, $D_{\mu}=\partial_{\mu}+i\,e\,A_{\mu}+i\,\frac{g}{2}\,\epsilon_{\mu\nu\alpha\beta}\, (k_{AF})^{\nu}\,F^{\alpha\beta}$, in the context of the Dirac equation. Here, $(k_{AF})^{\lambda}$ is the Carroll–Field–Jackiw (CFJ) four-vector, and $g$ is the non minimal coupling magnitude. In Ref. \cite{RC2}, author proposed a new CPT-even non minimal coupling between the fermionic and gauge fields in the context of the Dirac equation, where $D_{\mu}=\partial_{\mu}+i\,e\,A_{\mu}+i\,\frac{\lambda^{(e)}}{2}\, (K_{F})_{\mu\nu\alpha\beta}\,\gamma^{\nu}\,F^{\alpha\beta}$. Here $(K_{F})_{\mu\nu\alpha\beta}$ is the background tensor that governs the Lorentz violation in the CPT-even electrodynamics of the SME \cite{b3,b4,b6} with the same symmetries as that of the Riemann tensor in General Relativity, as well as zero on double-trace condition $(K_{F})^{\mu\nu}_{\mu\nu}=0$ \cite{cc2,b8}. The author in Ref. \cite{cc1} studied the spontaneous symmetry breaking and present a method to derive the gauge-field propagators. Also, they had given the bosonic field action by $S_{action}=-\frac{1}{4}\,\int d^{4}x\, (K_{F})_{\mu\nu\alpha\beta}\,F^{\mu\nu}\,F^{\alpha\beta}$, where $F^{\mu\nu}$ is the electromagnetic field tensor. The decomposition of the tensor field was given in details in Refs. \cite{RC2,cc1} and the current limits on the coefficients of Lorentz-violating were given in Ref. \cite{VAK2}.

Inspired by Refs. \cite{RC2,cc1}, the background of Lorentz symmetry violation and CPT-even non minimal coupling established by a tensor field in the context of the Klein-Gordon equation was proposed in Ref. \cite{aa15} where, the term $-\frac{g}{4}\,(K_{F})_{\mu\nu\alpha\beta}\,F^{\mu\nu}\,F^{\alpha\beta}$ is coupled through a substitution $p^{\mu}\,p_{\mu} \to p^{\mu}\,p_{\mu}-\frac{g}{4}\,(K_{F})_{\mu\nu\alpha\beta}\,F^{\mu\nu}\,F^{\alpha\beta}$. Here the tensor field $(K_{F})_{\mu\nu\alpha\beta}$ is same as stated above and is CPT even, {\it i.e.}, it does not violate the CPT symmetry. Though CPT violation implies violation of Lorentz invariance \cite{OWG}, the reverse is not necessarily true. After that many authors have been studied the effects of Lorentz-violating in the relativistic quantum system, such as, quantum motions of spin-$0$ scalar bosons in Refs. \cite{aa10,aa11,aa12,aa15,ff9,EPL}, Klein–Gordon and generalized Klein-Gordon oscillator in Refs. \cite{aa13,aa16,IJMPA,MPLA,EPL2,IJGMMP,RSPA,INJP,ZWL,EPL4}, dynamics of spin-$1/2$ particles Refs. \cite{aa14,aa17,aa18,aa19,ff2,ff8}, and the Dirac oscillator Ref. \cite{RLLV}. Note that the Lorentz symmetry breaking effects was studied in non-relativistic limit also, for example, studies on Aharonov–Bohm effect \cite{ss1}, in Rashba-type coupling \cite{ss2,ss3}, in Landau-type quantization \cite{ss4}, on geometric phases for a Dirac neutral particle \cite{ss5,ss6}, in a Dirac neutral particle inside a two-dimensional quantum ring \cite{ss7}, in a spin-orbit coupling for a neutral particle \cite{ss8}, on a Coulomb-like central potential induced by the LSV on the harmonic oscillator \cite{ss9}, neutral particle with a permanent magnetic dipole moment that interacts with an electric field \cite{ASO3}, non-relativistic quantum effects associated with a privileged direction in the space-time \cite{ASO4}, a neutral particle in the presence of an attractive inverse-square potential \cite{ASO2}.

In this contribution, we study the generalized Klein-Gordon oscillator under the effects of Lorentz-Violating. We choose a scenario of Lorentz symmetry breaking with the electromagnetic field configuration defined by a Cornell-type potential form electric field and a linear magnetic field. Then, we choose a Coulomb- and Cornell-type potential form function and determines the bound-state eigenvalue solutions of the generalized oscillator field. We analyze the effects of Lorentz symmetry breaking as well as electromagnetic field on the energy profile and the wave function and show that our results is different from those results obtained in Ref. \cite{ZWL} and gets modified.

This paper is summarized as follows: in {\tt section 2}, we study the Klein-Gordon field in a Lorentz symmetry violation environment and then introduce oscillator to the field. We solve this modified wave equation by choosing a Coulomb-type potential form function $f(r)=\frac{b}{r}$ ({\tt sub-section 2.1}) and a Cornell-type potential form function $f(r)=\Big(a\,r+\frac{b}{r}\Big)$; finally in {\tt section 3}, we present our results.  

\section{Analysis of Generalized Relativistic Quantum Oscillator Field Under Lorentz Symmetry Violation}

In this section, we study the relativistic quantum motions of a scalar field via the generalized Klein-Gordon oscillator under central potential effects induced by Lorentz symmetry breaking environment defined by a tensor field $(K_F)_{\mu\nu\alpha\beta}$ out of the SME. As stated above, this tensor field coupled with the electromagnetic field tensor $F^{\mu\nu}$ is introduced into the Klein-Gordon equation through a non-minimal substitution 
\begin{equation}
p^{\mu}\,p_{\mu} \to p^{\mu}\,p_{\mu}-\frac{g}{4}\,(K_F)_{\mu\nu\alpha\beta}\,F^{\mu\nu}\,F^{\alpha\beta}.
\label{sp}
\end{equation}
This tensor field $(K_F)_{\mu\nu\alpha\beta}$ can expressed by the parity-even sector given by $(\kappa_{DE})_{jk}=-2\,(K_F)_{0j0k}$, $(\kappa_{HB})_{jk}=\frac{1}{2}\,\epsilon_{jpq}\, \epsilon_{klm}\,(K_F)^{pqlm}$, and the parity-odd sector given by $(\kappa_{DB})_{jk}=-(\kappa_{HE})_{kj}= \epsilon_{kpq}\,(K_F)^{0jpq}$, respectively Refs. \cite{DC,DC2,b1,b3,b4,b2,b6,b7,RC2,cc1}. Next, for studies of the generalized Klein-Gordon oscillator, we replace the momentum operator $p_{\mu} \to \Big(p_{\mu}+i\,M\,\omega\,X_{\mu}\Big)$ Refs. \cite{aa13,aa16,IJMPA,MPLA,EPL3,RSPA,INJP,ZWL} and $p_{\mu}^{\,\dagger} \to \Big(p_{\mu}-i\,M\,\omega\,X_{\mu}\Big)$ in the modified equation (\ref{sp}), where $X_{\mu}=(0, f(r), 0, 0)= f (r)\,\delta^{r}_{\mu}$ is an arbitrary four-vector, and $\omega$ is the oscillation frequency.

Therefore, the relativistic quantum motions of scalar oscillator field via the generalized Klein-Gordon oscillator is described by the following wave equation \cite{IJMPA,MPLA,EPL4} ($c=1=\hbar$)
\begin{eqnarray}
&&\Bigg[-\Big(p^{\mu}+i\,M\,\omega\,X^{\mu}\Big)\,\Big(p_{\mu}-i\,M\,\omega\,X_{\mu}\Big)+\frac{g}{4}\,(K_F)_{\mu\nu\alpha\beta}\,F^{\mu\nu}\,F^{\alpha\beta}\Bigg]\,\Psi=M^2\,\Psi\nonumber\\
&&\Bigg[-\frac{\partial^2}{\partial\,t^2}-\Big(\vec{p}-i\,M\,\omega\,f(r)\,\hat{r}\Big)\bullet\Big(\vec{p}+i\,M\,\omega\,f(r)\,\hat{r}\Big)+\frac{g}{4}\,(K_F)_{\mu\nu\alpha\beta}\,F^{\mu\nu}\,F^{\alpha\beta}\Bigg]\,\Psi=M^2\,\Psi.
\label{1}
\end{eqnarray}

Using the properties of the tensor field $(K_F)_{\mu\nu\alpha\beta}$, one can write Eq. (\ref{1}) in flat space metric in cylindrical coordinates $(t, r, \phi, z)$ as
\begin{eqnarray}
&&\Bigg[-\frac{\partial^2}{\partial t^2}+\frac{1}{r}\,\Big(\frac{\partial}{\partial r}+M\,\omega\,f(r) \Big)\,\Big(r\,\frac{\partial}{\partial r}-M\,\omega\,r\,f(r)\Big) +\frac{1}{r^2}\, \frac{\partial^2}{\partial \phi^2}+\frac{\partial^2}{\partial z^2}\Bigg]\,\Psi\nonumber\\
&+&\Bigg[-\frac{g}{2}\,(\kappa_{DE})_{ij}\,E^{i}\,E^{j}+\frac{g}{2}\,(\kappa_{HB})_{ij}\,B^i\,B^j-g\,(\kappa_{DB})_{ij}\,E^i\,B^j\Bigg]\,\Psi=M^2\,\Psi.
\label{2}
\end{eqnarray}
That may be written as
\begin{eqnarray}
&&\Bigg[-\frac{\partial^2}{\partial t^2}+\frac{\partial^2}{\partial r^2}+ \frac{1}{r}\, \frac{\partial}{\partial r}-M\,\omega\,\Big(f'+\frac{f}{r}\Big)-M^2\,\omega^2\,f^2(r)+ \frac{1}{r^2}\,\frac{\partial^2}{\partial \phi^2}+\frac{\partial^2}{\partial z^2}\Bigg]\,\Psi\nonumber\\
&+&\Bigg[-\frac{g}{2}\,(\kappa_{DE})_{ij}\,E^{i}\,E^{j}+\frac{g}{2}\,(\kappa_{HB})_{ij}\,B^i\,B^j-g\,(\kappa_{DB})_{ij}\,E^i\,B^j-M^2\Bigg]\,\Psi=0.
\label{3}
\end{eqnarray}
Here prime denotes derivative w. r. t. the argument.

In this analysis, we consider the following electric field of the electromagnetic field configuration given by
\begin{equation}
\vec{E}=\vec{E}_1+\vec{E}_2,
\label{4}
\end{equation}
where the individual electric field is given by $\vec{E}_1=(E_1, 0, 0)$ with $E_1 \propto r \Rightarrow E_1=c_1\,r$ Refs.\cite{aa10,aa11,KB,CF,CF2,CF3,KB4,ME,IJMPA,EPL2,EPL4,RSPA,INJP}, where $c_1$ is a constant and $\vec{E_2}=(E_2, 0, 0)$ with $E_2 \propto \frac{1}{r} \Rightarrow E_2=\frac{c_2}{r}$ Refs. \cite{aa15,aa16,MPLA,EPL2,IJGMMP}, where $c_2$ is a constant. Therefore, we have the electric field $\vec{E}=\Big(c_1\,r+\frac{c_2}{r}\Big)\,\hat{r}$. Note that for $c_2 \to 0$, we have a linear radial electric field. This type of linear electric field is produced by a uniform volume distribution of the electric charges having volume charge density $\rho$. On the other hand, if $c_1 \to 0$, one will have a Coulomb-type radial electric field that is produced by uniform distribution of the linear charge density. This type of combined electric field along the radial direction is called the Cornell-type potential form electric field that has recently been studied in quantum systems in Ref. \cite{ZWL}.

Let us assume that there exists, inside an infinitely long non-conductor cylinder, a magnetization given by $\vec{M}=\kappa\,r\,\hat{z}$ (where $\kappa$ is a constant) \cite{DJG}. This example of magnetization is well known in the literature \cite{DJG,JACB,AHU}. It can be achieved by a distribution of non-conducting super-paramagnetic particles with an elliptical shape (single-domain) \cite{JACB}, like a core-at-shell coating system \cite{AHU}, where its major axis is parallel to the axis of the cylinder. The shape factor provides a larger demagnetization field between the adjacent particles in the radial direction (compared to the z-direction) as a pseudo-permanent magnet. The cylinder is constantly magnetized by an electromagnet with inverted conical poles. The conical shape of the poles yields a $\kappa$-coefficient to magnetization inside the sample. However, we should keep in mind that there exists an external magnetic field of the electromagnet. Since the cylinder is very long, this external magnetic field can be considered to be weak at the point of our analysis, thus, we can neglect it. Therefore, by following ref. \cite{DJG}, inside this non-conductor cylinder, this magnetization produces a magnetic field 
\begin{equation}
\vec{B}=\mu_0\,\kappa\,r\,\hat{z}=\chi\,r\,\hat{z}\quad,\quad \chi=\mu_0\,\kappa.
\label{mf}
\end{equation}
This type of magnetic field has been used in quantum system by various authors  \cite{aa12,CS,CS2,RSPA,INJP,EPL4}.
 
Based on the above crossed electric and magnetic field configuration, we consider a possible scenario of the Lorentz symmetry violation given by the following nonzero components 
\begin{equation}
(\kappa_{DE})_{11}=\kappa_1\quad,\quad (\kappa_{DB})_{13}=\kappa_2, 
\label{5}
\end{equation}
while $(\kappa_{HB})_{ij}=0$ of the tensor field $(K_F)_{\mu\nu\alpha\beta}$. Here $\kappa_1, \kappa_2$ are constants. 
 
Also by using the method of separation of the variables, one can always express the total wave function in terms of different variables. Let us choose an ansatz of the total wave function $\Psi (t, r, \phi, z)$ in terms of a radial wave function $\psi (r)$ as follows: 
\begin{equation}
\Psi (t, r, \phi, z)=e^{-i\,\varepsilon\,t}\,e^{i\,l\,\phi}\,e^{i\,k\,z}\,\psi (r),
\label{6}
\end{equation}
where $\varepsilon$ is energy of the scalar particle, $l=0,\pm 1,\pm 2,....$ are the eigenvalues of the $z$-component of the angular momentum operator, and $k$ is a constant.
 
We study below the generalized Klein-Gordon oscillator under a central potential effect induced by Lorentz symmetry breaking environment defined by the crossed electric and magnetic fields Eq. (\ref{4}) with the non-zero components Eq. (\ref{5}) of the tensor field. In this analysis, we have chosen two types of function, namely, a Coulomb-type which is inversely proportional to the inverse of the axial distance $(f \propto \frac{1}{r}\Rightarrow f(r)=\frac{b}{r})$, and a Cornell-type function the sum of linear and Coulomb-type function given by $f(r)=\Big(a\,r+\frac{b}{r}\Big)$ and analyzed the quantum mechanical systems.
 
\subsection{\bf Coulomb-Type Potential Form Function $f(r)=\frac{b}{r}$.}
 
In this section, we study the generalized Klein-Gordon oscillator by choosing a Coulomb-type potential form function $f(r)=\frac{b}{r}$. This type of function has been studied in quantum mechanical systems Refs. \cite{EPJP2,EPJP3,GERG,SZ,EPL4,IJMPA,MPLA}. 

Thereby, substituting (\ref{4})--(\ref{6}) and the Coulomb-type function into the Eq. (\ref{3}), we arrived at the following differential equation:
\begin{equation}
\psi''(r)+\frac{1}{r}\,\psi'(r)+\Big(\Pi-\frac{\tau^2}{r^2}-\Omega^2\,r^2\Big)\,\psi(r)=0,
\label{aa1}
\end{equation}
where we have defined different parameters 
\begin{equation}
\Pi=\varepsilon^2-M^2-k^2-g\,\kappa_1\,c_1\,c_2-g\,\kappa_2\,\chi\,c_2,\quad \Omega=\sqrt{\frac{g}{2}\,c^2_{1}\,\kappa_1+g\,c_1\,\chi\,\kappa_2}\,\quad \tau=\sqrt{l^2+M^2\,\omega^2\,b^2+\frac{g}{2}\,c^2_{2}\,\kappa_1}.
\label{aa2}
\end{equation}

Equation (\ref{aa1}) is a second-order differential equations which can be solved using an appropriate method or technique. In literature, several authors have been employed different techniques or methods to obtain the exact or approximate solutions of the wave equations, such as, power series method, factorization method, super-symmetric quantum mechanics, path integral method, the parametric Nikiforov- Uvarov (NU) and its function analysis method etc.. Of these, the parametric Nikiforov-Uvarov (NU) method \cite{AFN,MA} is of our particular interest which is based on hypergeometric-type second-order differential equation \cite{GS}. This NU method has been successfully applied in solving the quantum mechanical problems (see, Refs. \cite{GERG,EPJP2,EPJP3,EPL,EPL2,EPL3,EPL4,RSPA,INJP}).

To solve the above equation (\ref{aa1}), let us perform a change of variables via $s=\Omega\,r^2$ in the Eq. (\ref{aa1}), we have 
\begin{equation}
\psi''(s)+\frac{1}{s}\,\psi'(s)+\frac{1}{s^2}\,\Big(-\xi_1\,s^2+\xi_2\,s-\xi_3\Big)\,\psi (s)=0,
\label{aa3}
\end{equation}
where different parameters are
\begin{equation}
\xi_1=\frac{1}{4},\quad \xi_2=\frac{\Pi}{4\,\Omega},\quad \xi_3=\frac{\tau^2}{4}. 
\label{aa4}
\end{equation}

According to the parametric NU-method \cite{AFN,MA}, the wave functions of the following second-order differential equation
\begin{equation}
\frac{d^2 \psi (s)}{ds^2}+\frac{(\alpha_1-\alpha_2\,s)}{s\,(1-\alpha_3\,s)}\frac{d \psi (s)}{ds}+\frac{(-\xi_1\,s^2+\xi_2\,s-\xi_3)}{s^2\,(1-\alpha_3\,s)^2}\,\psi(s)=0
\label{A.1}
\end{equation}
are given by 
\begin{equation}
\psi (s)=s^{\alpha_{12}}(1-\alpha_3\,s)^{-\alpha_{12}-\frac{\alpha_{13}}{\alpha_3}}\, P^{(\alpha_{10}-1, \frac{\alpha_{11}}{\alpha_3}-\alpha_{10}-1)}_{n}(1-2\alpha_3\,s).
\label{A.2}
\end{equation}
And that the energy eigenvalues equation is given by
\begin{equation}
\alpha_2\,n-(2\,n+1)\,\alpha_5+(2\,n+1)\,(\sqrt{\alpha_9}+\alpha_3\,\sqrt{\alpha_8})+n\,(n-1)\,\alpha_3+\alpha_7+2\,\alpha_3\,\alpha_8+2\,\sqrt{\alpha_8\,\alpha_9}=0.
\label{A.3}
\end{equation}
The parameters $\alpha_4,\ldots,\alpha_{13}$ are obtained from the six parameters $\alpha_1,\ldots, \alpha_3$ and $\xi_1,\ldots,\xi_3$ as follows:
\begin{eqnarray}
&&\alpha_4=\frac{1}{2}\,(1-\alpha_1),\quad\alpha_5=\frac{1}{2}\,(\alpha_2-2\,\alpha_3),\quad \alpha_6=\alpha^2_{5}+\xi_1,\quad \alpha_7=2\,\alpha_4\,\alpha_{5}-\xi_2,\quad \alpha_8=\alpha^2_{4}+\xi_3,\nonumber\\
&&\alpha_9=\alpha_6+\alpha_3\,\alpha_7+\alpha^{2}_3\,\alpha_8,\quad \alpha_{10}=\alpha_1+2\,\alpha_4+2\,\sqrt{\alpha_8},\quad 
\alpha_{11}=\alpha_2-2\,\alpha_5+2\,(\sqrt{\alpha_9}+\alpha_3\,\sqrt{\alpha_8}),\nonumber\\
&&\alpha_{12}=\alpha_4+\sqrt{\alpha_8},\quad\alpha_{13}=\alpha_5-(\sqrt{\alpha_9}+\alpha_3\,\sqrt{\alpha_8}).
\label{A.4}
\end{eqnarray}

By Comparing Eq. (\ref{aa3}) with the Eq. (\ref{A.1}), we have
\begin{eqnarray}
&&\alpha_1=1,\quad \alpha_2=0,\quad \alpha_3=0,\quad \alpha_4=0,\quad \alpha_5=0,\quad \alpha_6=\xi_1,\quad \alpha_7=-\xi_2,\quad \alpha_8=\xi_3,\nonumber\\
&&\alpha_9=\xi_1,\quad \alpha_{10}=1+2\,\sqrt{\xi_3},\quad \alpha_{11}=2\,\sqrt{\xi_1},\quad \alpha_{12}=\sqrt{\xi_3},\quad \alpha_{13}=-\sqrt{\xi_1}.
\label{aa5}
\end{eqnarray}
Using $\xi_1,...,\xi_3$ and Eq. (\ref{aa5}) into the Eq. (\ref{A.3}), one can find the following eigenvalues of the quantum system
\begin{equation}
\varepsilon_{n, l, k}=\pm\,\sqrt{M^2+k^2+2\,\Big(2\,n+1+\sqrt{l^2+M^2\,\omega^2\,b^2+\frac{g}{2}\,c^2_{2}\,\kappa_1}\Big)\,\sqrt{\frac{g}{2}\,c^2_{1}\,\kappa_1+g\,c_1\,\chi\,\kappa_2}+g\,\kappa_1\,c_1\,c_2+g\,\kappa_2\,\chi\,c_2},
\label{aa6}
\end{equation}
where $n=0,1,2,...$.

The normalized radial wave function is given by
\begin{equation}
\psi_{n,l,k} (s)=N_{n,l}\,s^{\frac{\tau}{2}}\,e^{-\frac{s}{2}}\,L^{(\tau)}_{n} (s).
\label{aa7}
\end{equation}
In terms of $r$ where $s=\Omega\,r^2$, the radial wave function becomes
\begin{equation}
\psi_{n,l,k} (r)=N_{n,l}\,(\Omega)^{\frac{\tau}{2}}\,r^{\tau}\,e^{-\frac{1}{2}\,\Omega\,r^2}\,L^{(\tau)}_{n} (\Omega\,r^2),
\label{aa8}
\end{equation}
$N_{n,l}$ is the normalization constant which can be determined by the following condition
\begin{equation}
\int^{\infty}_{0}\,r\,dr\,|\psi (r)|^2=1.
\label{aa9}
\end{equation}
Thereby, substituting (\ref{aa8}) in the condition (\ref{aa9}), we have the normalization constant
\begin{equation}
N_{n,l}=(\Omega)^{\frac{1}{2}}\,\sqrt{\frac{2\,(n!)}{(n+\tau)!}},
\label{aa10}
\end{equation}
where $\tau$ is given in (\ref{aa2}) and we have used the following relation of the generalized Laguerre polynomials which is orthogonal over the range $(0, \infty]$ w. r. t. weighting function $s^{\tau}\,e^{-s}$ given by
\begin{equation}
\int^{\infty}_{0} s^{\tau}\,e^{-s}\,L^{(\tau)}_{n}\,(s)\,L^{(\tau)}_{m}\,(s)\,ds=\left (\frac{(n+\tau)!}{n!}\right)\,\delta_{n\,m}.
\label{aa11}
\end{equation}

Thus, one can write the normalized wave function from (\ref{aa7}) and using (\ref{aa10}) becomes:
\begin{equation}
\psi_{n, l, k} (s)=(\Omega)^{\frac{1}{2}}\left(\frac{2\,(n!)}{\Big(n+\sqrt{l^2+M^2\,\omega^2\,b^2+\frac{g}{2}\,c^2_{2}\,\kappa_1}\Big)!}\right)^{\frac{1}{2}}\,s^{\frac{\sqrt{l^2+M^2\,\omega^2\,b^2+\frac{g}{2}\,c^2_{2}\,\kappa_1}}{2}}\,e^{-\frac{s}{2}}\,L^{\Big(\sqrt{l^2+M^2\,\omega^2\,b^2+\frac{g}{2}\,c^2_{2}\,\kappa_1}\Big)}_{n}\,(s),
\label{aa12}
\end{equation}

Equation (\ref{aa6}) is the relativistic energy eigenvalues and Eq. (\ref{aa7}) is the normalized radial wave function of a oscillator field by choosing a Coulomb-type function under the Lorentz symmetry breaking environment. We can see that the energy eigenvalue and the wave function are influenced by the Lorentz symmetry breaking effects defined by the parameters $\Big\{g, \kappa_1, \kappa_2 \Big\}$ as well as the electromagnetic field parameters $\Big\{c_1, c_2, \chi \Big\}$ in comparison to the result without Lorentz symmetry breaking effects. This result without the effects of Lorentz symmetry violation can be obtained by setting the parameters $\kappa_1=0=\kappa_2$ in the radial Eq. (\ref{aa1}) which becomes the Bessel's second-order differential equation whose solutions are well-known.

\subsection{\bf Cornell-Type potential function $f(r)=\Big(a\,r+\frac{b}{r}\Big)$. }
 
In this section, we study the generalized Klein-Gordon oscillator under Lorentz-Violating effects by choosing a Cornell-type potential form function $f(r)=\Big(a\,r+\frac{b}{r}\Big)$. The Cornell-type potential which consist of both short and long-ranges interactions has been studied in different branches of physics. In Refs. \cite{EPJP2,EPJP3,GERG,SZ,AA1,AA2,AA3,AA4}, authors have used this type function for the studies of the generalized Klein-Gordon or the Dirac oscillator in quantum systems. 

Thereby, substituting (\ref{4})--(\ref{6}) and the Cornell-type function into the Eq. (\ref{3}), we arrive at the following differential equation:
\begin{equation}
\psi '' (r)+\frac{1}{r}\,\psi '(r)+\left(\Lambda-\frac{\tau^2}{r^2}-\delta^2\,r^2 \right)\,\psi (r)=0,
\label{bb1}
\end{equation}
where $\tau$ is defined earlier and 
\begin{equation}
\Lambda=\Pi-2\,M\,\omega\,a-2\,a\,b\,M^2\,\omega^2,\quad \delta=\sqrt{M^2\,\omega^2\,a^2+\Omega^2}.
\label{bb2}
\end{equation}
Performing a change of variables via $s=\delta\,r^2$ into the Eq. (\ref{bb1}), one can obtained the following second-order differential equation
\begin{equation}
\psi''(s) +\frac{1}{s}\,\psi'(s)+\frac{1}{s^2}\,\Big(-\zeta_1\,s^2+\zeta_2\,s-\zeta_3\Big)\,\psi(s)=0,
\label{bb3}
\end{equation}
where 
\begin{equation}
\zeta_1=\frac{1}{4}\quad,\quad \zeta_2=\frac{\Lambda}{4\,\delta}\quad,\quad \zeta_3=\frac{\tau^2}{4}. 
\label{bb4}
\end{equation}

By Comparing Eq. (\ref{bb3}) with the Eq. (\ref{A.1}), we have
\begin{eqnarray}
&&\alpha_1=1,\quad \alpha_2=0,\quad \alpha_3=0,\quad \alpha_4=0,\quad \alpha_5=0,\quad \alpha_6=\zeta_1,\quad \alpha_7=-\zeta_2,\quad \alpha_8=\zeta_3,\nonumber\\
&&\alpha_9=\zeta_1,\quad \alpha_{10}=1+2\,\sqrt{\zeta_3},\quad \alpha_{11}=2\,\sqrt{\zeta_1},\quad \alpha_{12}=\sqrt{\zeta_3},\quad \alpha_{13}=-\sqrt{\zeta_1}.
\label{bb7}
\end{eqnarray}
Using $\zeta_1,...,\zeta_3$ and Eq. (\ref{bb7}) into the Eq. (\ref{A.3}), one can find the following eigenvalues of the quantum system
\begin{equation}
\varepsilon_{n,l,k}=\pm\,\sqrt{M^2+k^2+2\,M\,\omega\,a+2\,a\,b\,M^2\,\omega^2+g\,\kappa_1\,c_1\,c_2+g\,\kappa_2\,\chi\,c_2+2\,\Big(2\,n+1+\tau\Big)\,\sqrt{M^2\,\omega^2\,a^2+\Omega^2}},
\label{bb5}
\end{equation}
where $n=0,1,2,...$ and $\tau$ is given in (\ref{aa2}).

The normalized radial wave functions are given by
\begin{equation}
\psi_{n,l,k} (s)=(\delta)^{\frac{1}{2}}\,\left (\frac{2\,(n!)}{\Big(n+\sqrt{l^2+M^2\,\omega^2\,b^2+\frac{g}{2}\,c^2_{2}\,\kappa_1}\Big)!}\right)^{\frac{1}{2}}\, s^{\frac{\sqrt{l^2+M^2\,\omega^2\,b^2+\frac{g}{2}\,c^2_{2}\,\kappa_1}}{2}}\,e^{-\frac{s}{2}}\,L^{(\sqrt{l^2+M^2\,\omega^2\,b^2+\frac{g}{2}\,c^2_{2}\,\kappa_1})}_{n}\,(s),
\label{bb6}
\end{equation}
where $\delta$ is given in Eq. (\ref{bb2}) and $\Omega$ in Eq. (\ref{aa2}).

Equation (\ref{bb5}) is the relativistic energy eigenvalues and Eq. (\ref{bb6}) is the normalized radial wave function of a oscillator field by choosing a Cornell-type function under Lorentz symmetry breaking environment. We can see that the energy eigenvalues and the wave function are influenced by Lorentz symmetry breaking defined by the parameters $\{g, \kappa_1, \kappa_2\}$, the electromagnetic field parameters $\{c_1, c_2, \chi\}$, and the Cornell-type function $f(r)$ with parameter $\{a, b\}$ and gets modified in comparison to the results obtained in the previous section as well as without Lorentz symmetry breaking effects. The result without the effects of Lorentz symmetry violation can be obtained by setting the parameters $\kappa_1=0=\kappa_2$ in the radial Eq. (\ref{bb1}) which becomes the Bessel's differential equation whose solutions are well-known. Furthermore, for zero magnetic field, $\chi \to 0$, one can show that the energy eigenvalue (\ref{bb5}) reduces to the result obtained in Ref. \cite{ZWL} (see Eq. (23) in Ref. \cite{ZWL}). Thus, the presence of a linear magnetic field in quantum system increases the energy levels of a oscillator field in comparison to the eigenvalue Eq. (23) obtained in Ref. \cite{ZWL}.

\section{Conclusions}

In this work, we have studied the generalized Klein-Gordon oscillator under Lorentz symmetry violation effects defined by a tensor $(K_F)_{\mu\nu\alpha\beta}$ out of the SME. The Lorentz symmetry violation is introduced by this tensor field $(K_F)_{\mu\nu\alpha\beta}$ through a non-minimal substitution and the scenario of this violation is chosen by the non-zero components $(\kappa_{DE})_{11}=\kappa_1, (\kappa_{DB})_{13}=\kappa_2$ while $(\kappa_{HB})_{ij}=0$ of the tensor field. Furthermore, the electromagnetic field configuration is chosen by a Cornell-type radial electric field $\vec{E}=\Big(c_1\,r+\frac{c_2}{r}\Big)\,\hat{r}$ and a linear magnetic field $\vec{B}=\chi\,r\,\hat{z}$, respectively that contributes a central potential effect in quantum motions of scalar field. The generalized Klein-Gordon oscillator is examined by replacing the radial momentum operator $-i\,\vec{\nabla} \to \Big(-i\,\vec{\nabla}-i\,M\,\omega\,f(r)\,\hat{r}\Big)$ in the modified Klein-Gordon equation. 

For studies of the generalized Klein-Gordon oscillator, in {\tt sub-section 2.1} we have chosen a Coulomb-type potential form function $f(r)=\frac{b}{r}$ and derived the radial wave equation of the generalized Klein-Gordon oscillator which is a second-order differential equation. Using the well-known Nikiforov-Uvarov (NU) method, one can find the energy eigenvalues given by the Eq. (\ref{aa6}) and the normalized wave function given by the Eq. (\ref{aa7}) of the oscillator fields. We have seen that the Lorentz symmetry violation given by the parameters $\Big\{g, c_1, c_2, \kappa_1, \kappa_2, \chi\Big\}$ modified the energy spectrum and the wave function of the oscillator fields. In {\tt sub-section 2.2}, we have chosen another function called the Cornell-type potential form function given by $f(r)=\Big(a\,r+\frac{b}{r}\Big)$ and derived the radial wave equation of the generalized Klein-Gordon oscillator. Then using the same Nikiforov-Uvarov method, we have obtained the energy eigenvalue given by the Eq. (\ref{bb5}) and the normalized wave function given by the Eq. (\ref{bb6}). We have seen that the Lorentz symmetry violation given by the parameters $\Big\{g, c_1, c_2, \kappa_1, \kappa_2, \chi\Big\}$ and the Cornell-type function $f(r)$ modified the energy eigenvalues and the wave function of the oscillator fields.

Here, we studied a relativistic quantum oscillator model via the generalized Klein-Gordon oscillator under Lorentz-Violating effects defined by a tensor field. We determined the manner in which Lorentz symmetry violation modified the energy profile and the wave function of these oscillator field. We have seen that the energy spectrum gets modified in comparison to the Standard Landau levels (standard corresponds to without Lorentz symmetry violation effects). From the first experimental realization of the one-dimensional Dirac oscillator Ref. \cite{JAFV}, researchers have interested on the Klein-Gordon oscillator which has many applications in different branches of physics. The presented perturbation in the eigenvalues may be useful for simulation of a series of physical systems, for instance, vibrational spectrum of diatomic molecules \cite{SMI}, binding of heavy quarks \cite{CQ,MC}, quark–antiquark interaction \cite{EE}. From the observational point of view, it is clear that to have an observable modification in the energy eigenvalues, huge number of particles in the state is needed, otherwise the magnitude of the effect to a real spectrum may not be strong enough to be observed.

\section*{Acknowledgement}

We sincerely acknowledged the anonymous kind referee(s) for his/her valuable comments and suggestions.

\section*{Conflict of Interest}

There is no conflict of interest regarding publication of this paper.

\section*{Data Availability}

No new data are generated in this paper.

\section*{Funding Statement}

No fund has received for this paper.


\begin{thebibliography}{99}


\bibitem{DC3} V. A. Kostelecky and R. Potting, Phys. Rev. {\bf D 51}, 3923 (1995).

\bibitem{DC4} V. A. Kostelecky and R. Potting,  Phys. Lett. {\bf B 381}, 89 (1996).

\bibitem{DC} D. Colladay and V. A. Kostelecky, Phys. Rev. {\bf D 55}, 6760 (1997).

\bibitem{DC2} D. Colladay and V. A. Kostelecky, Phys. Rev. {\bf D 58}, 116002 (1998).

\bibitem{DC0} V. Alan Kostelecky, Phys. Rev. {\bf D 69}, 105009 (2004).

\bibitem{b1} S. Carroll, G. Field and R. Jackiw, Phys. Rev. {\bf D 41}, 1231 (1990).

\bibitem{b3} V. A. Kostelecky and M. Mewes, Phys. Rev. Lett. {\bf 87}, 251304 (2001).

\bibitem{b4} V. A. Kostelecky and M. Mewes, Phys. Rev. {\bf D 66}, 056005 (2002).

\bibitem{b2} A. P. B. Scarpelli, H. Belich, J. L. Boldo, L. P. Colatto, J. A. Helayel-Neto and A. L. M. A. Nogueira, Nucl. Phys. B (Proc. Suppl.) {\bf 127}, 105 (2004).

\bibitem{b6} V. A. Kostelecky and M. Mewes, Phys. Rev. Lett. {\bf 97}, 140401 (2006).

\bibitem{b7} V. A. Kostelecky and M. Mewes, Phys. Rev. {\bf D 80}, 015020 (2009).

\bibitem{RC2} R. Casana, M. M. Ferreira Jr., E. Passos, F. E. P. dos Santos and E. O. Silva, Phys. Rev. {\bf D 87}, 047701 (2013).

\bibitem{cc2} Q. G. Bailey and V. A. Kostelecky, Phys. Rev. {\bf D 70}, 076006 (2004). 

\bibitem{b8} G. Betschart, E. Kant, and F. R. Klinkhamer, Nucl. Phys. {\bf B 815}, 198 (2009).

\bibitem{cc1} H. Belich, F. J. L. Leal, H. L. C. Louzada and M. T. D. Orlando, Phys. Rev. {\bf D 86}, 125037 (2012).

\bibitem{VAK2} V. A. Kostelecky and N. Russell, Rev. Mod. Phys. {\bf 83}, 11 (2011). 

\bibitem{aa15} K. Bakke and H. Belich, Ann. Phys. (N. Y.) {\bf 360}, 596 (2015).

\bibitem{OWG} O. W. Greenberg, Phys. Rev. Lett. {\bf 89}, 231602 (2002).

\bibitem{aa10} K. Bakke and H. Belich, Ann. Phys. (N. Y.) {\bf 373}, 115 (2016).

\bibitem{aa11} R. L. L. Vitoria, H. Belich and K. Bakke, Adv. High Energy Phys. {\bf 2017}, 6893084 (2017).

\bibitem{aa12} R. L. L. Vitoria, K. Bakke and H. Belich, Ann. Phys. (N. Y.) {\bf 399}, 117 (2018).

\bibitem{ff9} R. L. L. Vitoria and H Belich, Eur. Phys. J. {\bf D 75}, 291 (2021).

\bibitem{EPL} F. Ahmed, EPL {\bf 136}, 41002 (2021).

\bibitem{aa13} R. L. L. Vitoria and H. Belich, Eur. Phys. J. C {\bf 78}, 999 (2018).

\bibitem{aa16} R. L. L. Vitoria, H. Belich and K. Bakke, Eur. Phys. J. Plus {\bf 132}, 25 (2017).

\bibitem{IJMPA} F. Ahmed, Int. J. Mod. Phys. {\bf A 36}, 2150128 (2021).

\bibitem{MPLA} F. Ahmed, Mod. Phys. Lett. {\bf A 36}, 2150274 (2021).

\bibitem{EPL2} F. Ahmed, EPL {\bf 136}, 61001 (2021).

\bibitem{IJGMMP} F. Ahmed, Int. J. Geom. Meths. Mod. Phys. {\bf 19}, 2150059 (2022).

\bibitem{RSPA} F. Ahmed, Proc. Roy. Soc. {\bf A 478}, 20220091 (2022).

\bibitem{INJP} F. Ahmed, Indian J. Phys. (2022), https://doi.org/10.1007/s12648-022-02438-5.

\bibitem{EPL4} F. Ahmed, EPL {\bf 139}, 30001 (2022).

\bibitem{ZWL} E.-Q. Wang, H. Chen, Y. Yang,  Z.-W. Long and H. Hassanabadi, Acta Physica Sinica {\bf 71} (6), 060301 (2022).

\bibitem{aa14} K. Bakke and H. Belich, Ann. Phys. (N. Y.) {\bf 333}, 272 (2013).

\bibitem{aa17} K. Bakke and H. Belich, Int. J. Mod. Phys. {\bf A 30}, 1550197 (2015).

\bibitem{aa18} A. G. de Lima, H. Belich and K. Bakke, Ann. Phys. (Berlin) {\bf 526}, 514 (2014).

\bibitem{aa19} K. Bakke, H. Belich and E. O. Silva, Ann. Phys. (Berlin) {\bf 523}, 910 (2011).

\bibitem{ff2} A. S. Oliveira, K. Bakke and H Belich, Int. J. Theor. Phys. {\bf 59}, 3396 (2020).

\bibitem{ff8} K. Bakke and H Belich, Int. J. Mod. Phys. {\bf A 34}, 1950116 (2019).

\bibitem{RLLV} R. L. L. Vitoria and H. Belich, Eur. Phys. J. Plus {\bf 135}, 247 (2020).

\bibitem{ss1} H. Belich, E. O. Silva, M. M. Ferreira Jr., M. T. D. Orlando, Phys. Rev. {\bf D 83}, 125025 (2011).

\bibitem{ss2} K. Bakke and H. Belich, Ann. Phys. (Leipzig) {\bf 526}, 1 (2013).

\bibitem{ss3} K. Bakke and H. Belich, Ann. Phys. (N. Y.) {\bf 354}, 1 (2015).

\bibitem{ss4} K. Bakke, H. Belich, J. Phys. G Nucl. Part. Phys. {\bf 42}, 095001 (2015)

\bibitem{ss5} K. Bakke and H. Belich, J. Phys. G Nucl. Part. Phys. {\bf 39}, 085001 (2012).

\bibitem{ss6} K. Bakke, E. O. Silva and H. Belich, J. Phys. G Nucl. Part. Phys. {\bf 39}, 055004 (2012).

\bibitem{ss7} K. Bakke and H. Belich, J. Phys. G Nucl. Part. Phys. {\bf 40}, 065002 (2013).

\bibitem{ss8} H. Belich, K. Bakke, Int. J. Mod. Phys. {\bf A 30}, 1550136 (2015).

\bibitem{ss9} K. Bakke and H. Belich, Eur. Phys. J. Plus {\bf 127}, 102 (2012).

\bibitem{ASO3} K Bakke and H Belich, Commun. Theor. Phys. {\bf 72}, 105204 (2020).

\bibitem{ASO4} A. S. Oliveira, K. Bakke and H. Belich, Eur. Phys. J. D {\bf 76}, 36 (2022). 

\bibitem{ASO2} K. Bakke and H. Belich, Int. J. Mod. Phys. {\bf A 36}, 2150067 (2021).

\bibitem{EPL3} F. Ahmed, EPL {\bf 130}, 40003 (2020).

\bibitem{KB} K. Bakke and H. Belich, Ann. Phys. (N. Y.) {\bf 354}, 1 (2015).

\bibitem{CF} C. Furtado, J. R. Nascimento and L. R. Ribeiro, Phys. Lett. {\bf A 358}, 336 (2006).

\bibitem{CF2} L. R. Ribeiro, C. Furtado and J. R. Nascimento, Phys. Lett. {\bf A 348}, 135 (2006).

\bibitem{CF3} K. Bakke, L. R. Ribeiro and C. Furtado, Cent. Euro. J. Phys. {\bf 8}, 893 (2010).

\bibitem{KB4} K. Bakke and H. Belich, Eur. Phys. J. Plus {\bf 127}, 102 (2012).

\bibitem{ME} M. Ericsson and E. Sj\"{o}qvist, Phys. Rev. {\bf A 65}, 013607 (2001).

\bibitem{DJG} D. J. Griffiths, {\tt Introduction to Electrodynamics}, Prentice Hall, New Jersey
(1999).

\bibitem{JACB} J. A. C. Bland and B. Heinrich, {\tt Ultrathin Magnetic Structures IV}, Springer-Verlag, Berlin, Germany (1994).

\bibitem{AHU} A. H. Lu, E. L. Salabas and F. Schuth, Angew. Chem. Int. Ed. {\bf 46}, 1222 (2007).

\bibitem{CS2} K. Bakke, C. Salavador and H. Belich, Int. J. Mod. Phys. {\bf A 33}, 1850216 (2018).

\bibitem{CS} K. Bakke and C. Salavador, Eur. Phys. J. Plus {\bf 133}, 123 (2018).

\bibitem{GERG} F. Ahmed, Gen. Relativ. Gravit. {\bf 51}, 69 (2019).

\bibitem{EPJP2} Marc de Montigny, H. Hassanabadi, J. Pinfold and S. Zare, Eur. Phys. J. Plus {\bf 136}, 788 (2021).

\bibitem{EPJP3} Marc de Montigny, J. Pinfold, S. Zare and H. Hassanabadi, Eur. Phys. J. Plus {\bf 137}, 54 (2022).

\bibitem{SZ} S. Zare, H. Hassanabadi and M. de Montigny, Gen. Relativ. Gravit. {\bf 52}, 25 (2020).

\bibitem{AFN} A. F. Nikiforov and V. B. Uvarov, {\tt Special functions of mathematical physics}, Birkhauser, Basel (1988).

\bibitem{MA} M. Abramowitz and I. A. Stegum, {\tt Hand book of Mathematical Functions}, Dover Publications Inc., New York (1965). 

\bibitem{GS} G. Sezgo, {\tt Orthogonal Polynomials}, American Mathematical Society, New York (1939).

\bibitem{AA1} F. Ahmed, Adv. High Energy Phys. {\bf 2020}, 8107025 (2020).

\bibitem{AA2} F. Ahmed, Eur. Phys. J. C {\bf 80}, 211 (2020).

\bibitem{AA3} F. Ahmed, Sci. Rep. {\bf 11}, 1742 (2021).

\bibitem{AA4} B. C. Lutfuoglu, J. Kriz, P. Sedaghatnia and H. Hassanabadi, Eur. Phys. J. Plus {\bf 135}, 691 (2020).

\bibitem{JAFV} J. A. F. -Villafane, E. Sadurni, S. Barkhofen, U. Kuhl, F. Mortessagne and T. H. Seligman, Phys. Rev. Lett. {\bf 111}, 170405 (2013).

\bibitem{SMI} S. M. Ikhdair, Int. J. Mod. Phys. {\bf C 20}, 1563 (2009)

\bibitem{CQ} C. Quigg and J. L. Rosner, Phys. Rep. {\bf 56}, 167 (1979).

\bibitem{MC} M. Chaichian and R. Kogerler, Ann. Phys. (N. Y.) {\bf 124}, 61 (1980).

\bibitem{EE} E. Eichten et al., Phys. Rev. Lett. {\bf 35}, 369 (1975).

\end{thebibliography}
\end{document}